# Doping-less Tunnel Field Effect Transistor: Design and Investigation

M. Jagadesh Kumar, *Senior Member, IEEE* and Sindhu Janardhanan

*Abstract*— Using calibrated simulations, we report a detailed study of the doping-less tunnel field effect transistor (TFET) on a thin intrinsic silicon film using charge plasma concept. Without the need for any doping, the source and drain regions are formed using the charge plasma concept by choosing appropriate work functions for the source and drain metal electrodes. Our results show that the performance of the doping-less TFET is similar to that of a corresponding doped TFET. The doping-less TFET is expected to be free from problems associated with random dopant fluctuations. Further, fabrication of doping-less TFET does not require high-temperature doping/annealing processes and therefore, cuts down the thermal budget opening up the possibilities for fabricating TFETs on single crystal silicon-on-glass substrates formed by wafer scale epitaxial transfer.

## I. INTRODUCTION

Tunnel field-effect transistors (TFET) are attracting wide attention because of their low subthreshold swing and low OFF-state leakage current [1-8]. Since the channel current is controlled by the tunneling mechanism on the source side, TFETs are more immune to short-channel effects (such as $V_T$ roll-off) unlike the conventional nanoscale MOSFETs [9-11]. However, the low ON-state current in Si TFETs, due to poor band-to-band tunneling efficiency, is a major challenge to be overcome. This problem is being extensively studied using strain, hetero-structures, low bandgap materials, high-k gate insulators and nanowires [6, 11-14]. The other problem with the TFET is that in aggressively scaled devices, random variability in transistor performance due to random dopant fluctuations (RDF) can become significant [15]. The effects of RDF, such as an unacceptably large increase in the OFF-state current, have recently been demonstrated in TFETs [16-19]. The presence of doped source and drain regions in TFETs also necessitates a complex thermal budget due to the need for ion implantation and expensive thermal annealing techniques [20-22]. Abrupt junctions are essential for efficient tunneling in TFETs [1,2,7]. However, creating abrupt junctions using high temperature processes is not easy due to the diffusion of the dopant atoms from the source/drain regions into the channel.

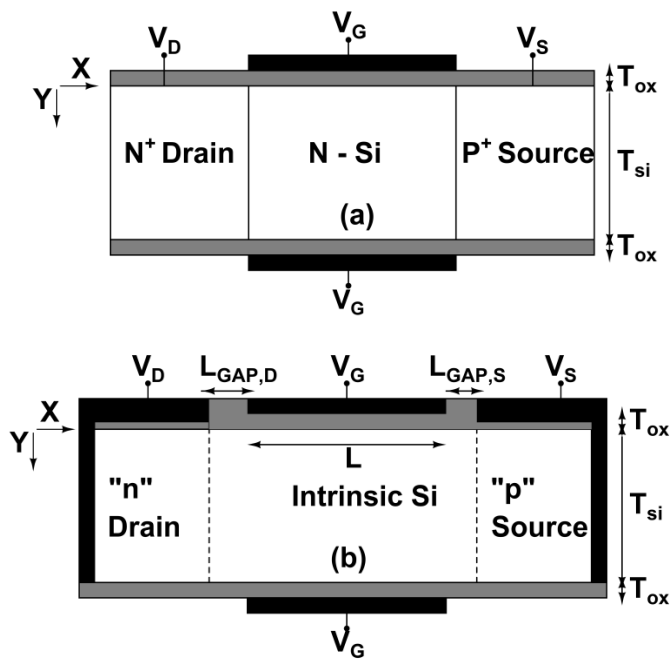

Fig. 1. Cross sectional view of (a) the conventional double gate TFET and (b) the doping-less TFET.

In this work, we report a detailed study on the doping-less TFET on intrinsic silicon film using the charge plasma concept [23-26]. In the doping-less TFET structure, without the need for any doping, the "p" source and the "n" drain are induced in the intrinsic silicon body by choosing the source and drain metal electrodes with suitable work functions. On the basis of calibrated simulation results, we demonstrate that the doping-less TFET's performance is similar to that of a corresponding doped TFET. Our idea can also be extended to realize the electron-hole bilayer TFET on an intrinsic semiconductor layer [27] but without the need for doping the source and drain regions. It may be pointed out that the doping-less TFET is not a combination of Schottky barrier FET (SBFET) and TFET. This is because the doping-less TFET does not have a Schottky junction between the induced "p" source and the channel. However, in the SBFET, the Schottky junction is between the metal source and the channel [28].

The doping-less TFET has the potential to realize TFETs using a low temperature process on single crystal silicon-on-glass substrates [29] as ion-implantation and subsequent high-temperature annealing are not required to fabricate the doping-less TFET. A possible realization of TFETs without a high

This work was supported in part by NXP (Philips) Chair Professorship awarded to M. J. Kumar. The authors are from the Department of Electrical Engineering, Indian Institute of Technology, Delhi, Hauz Khas, New Delhi 110 016, India.

Email: mamidala@ee.iitd.ac.in; sindhuramaswamy@gmail.com

temperature budget on silicon-on-glass substrates not only makes them suitable for electronics such as displays but also leads to bio- and opto-compatibility [30]. Since a charge plasma p-n junction has already been experimentally demonstrated [23] and the charge plasma based doping-less bipolar junction transistor [25,26] and the junction-less TFET have been reported [8], we believe that our results may provide the incentive for further exploration of the doping-less TFET.

## II. DEVICE STRUCTURE AND SIMULATION PARAMETERS

The cross sectional views of the conventional doped TFET and the doping-less TFET are shown in Fig. 1. The parameters for the conventional double-gate TFET used in our simulation are [6]: channel region doping $N_D = 1 \times 10^{17}/cm^3$, $P^+$ source doping $N_A = 1 \times 10^{20}/cm^3$, $N^+$ drain doping $N_D = 5 \times 10^{18}/cm^3$, silicon thickness ($T_{si}$) = 10nm, gate oxide thickness ($T_{ox}$) = 3nm, channel length (L) = 50 nm and gate work function = 4.5 eV.

The simulation parameters for the doping-less TFET are same as above except for the formation of the "p" source and "n" drain on the intrinsic silicon body with an intrinsic carrier concentration $n_i = 1.0 \times 10^{15}/cm^3$. To maintain uniform induced carrier distribution throughout the silicon thickness in the source and drain regions, from the source-Si interface to the Si-buried oxide interface along the Y-direction, the silicon film thickness has to be kept within the Debye length, i.e., $L_D = \sqrt{(\varepsilon_{si} V_T)/(qN)}$ where $\varepsilon_{si}$ is the dielectric constant of silicon, $V_T$ is the thermal voltage, and N is the carrier concentration in the body [24]. Our results do not change even if the silicon body carrier concentration is higher than $n_i$ (up to $10^{17}/cm^3$) due to any unintentional doping of the silicon body which is normally the case when the silicon films are grown epitaxially. We have chosen a film thickness of 10 nm as used in other simulation based TFET works [6,11,31,32]. Since we have not considered any quantum mechanical effects in our study, we have limited the choice of the silicon film thickness to 10 nm. For silicon film thicknesses below 10 nm, one would have to consider quantum mechanical effects [33].

In the doping-less TFET, the "p" source and "n" drain regions are formed using the charge plasma concept [23-26]. Under thermal equilibrium conditions, for creating the "n" drain region by inducing electrons with a concentration similar to the $N^+$ drain doping of the reference device in the intrinsic silicon body, hafnium (work function=3.9 eV) is employed as the drain metal electrode. Similarly, for creating the "p" source region by inducing holes with a concentration similar to the $P^+$ source doping of the reference device in the intrinsic silicon body, platinum (work function = 5.93 eV [34]) is employed as the source metal electrode. We have inserted a 0.5 nm thick silicon dioxide between the source metal electrode and the silicon film to avoid the possibility of silicide formation. To minimize the chances of silicide formation, one could also use a thicker high-κ dielectric (for an EOT = 0.5 nm) between the source metal electrode and silicon. However, while choosing the dielectric thickness, care must be taken to make sure that the induced source carrier concentration is reasonably high. This will ensure that the subthreshold swing is not adversely affected. Silicide formation can also be suppressed using substrate bias during the metal sputtering [35]. We have inserted a 3.0 nm thick silicon dioxide between the drain metal electrode and silicon to induce the drain region electron concentration similar to that of the reference TFET [6]. The spacer oxide thickness between the source and gate electrodes ($L_{GAP,S}$) is kept at 3 nm and between the drain and gate electrodes ($L_{GAP,D}$) at 15 nm which is approximately equal to the depletion region width on the drain-channel side in a conventional doped TFET.

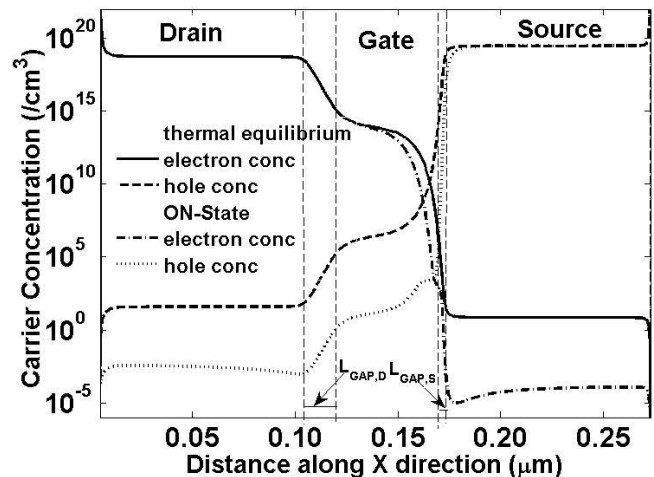

Fig. 2. The electron and hole concentrations of the doping-less TFET at 1 nm below the Si-SiO$_2$ interface under thermal equilibrium and ON-state conditions ($V_{GS} = V_{DS} = 1.0$ V).

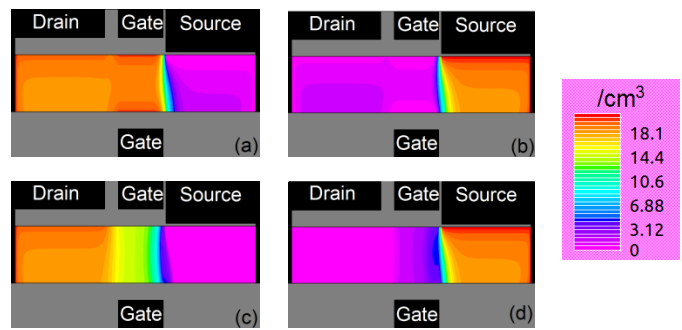

Fig. 3. Contour plots of (a) the electron and (b) the hole carrier concentrations for $V_{DS} = 100$ mV and (c) the electron and (d) the hole carrier concentrations for $V_{DS} = 1.0$ V with $V_{GS}$ fixed at 1.0 V.

The TFET structures in Fig. 1 are simulated using Silvaco Atlas V5.18.3.R device simulation tool [36]. Drift-diffusion current transport model is considered in the simulations. Lombardi mobility model and concentration dependent SRH recombination model are used [36]. We have not considered the charge induced band-gap narrowing (BGN) [37] while simulating the doping-less TFET. This is because our simulation tool does not have an appropriate model for the carrier induced BGN. Therefore, for a fair comparison, we have not included the dopant induced BGN effects in the conventional TFET. Non local band to band tunneling (BTBT) model is used to account for the spatial profile of the energy bands and also to account for the spatial separation of electrons generated in the conduction band from the holes generated in the valence band [6]. The grid points in the device are kept at 0.5 nm spacing in the y-direction across the thickness of the silicon film and at 0.2 nm spacing in the x-direction from the source-channel junction to the channel-drain junction. The results of the work reported in [6] are first reproduced to calibrate the model parameters as done in our earlier works [11,31,32] to avoid the use of default parameters of the simulator.

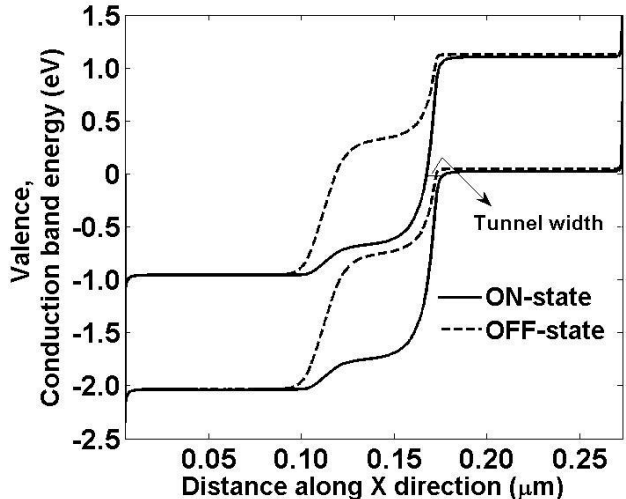

Fig. 4. Valence and conduction band energy in the OFF-state ($V_{GS}$ = 0 V, $V_{DS}$ = 1.0 V) and ON-state ($V_{GS}$ = $V_{DS}$ = 1.0 V) at 1 nm below the Si-SiO$_2$ interface for the doping-less TFET.

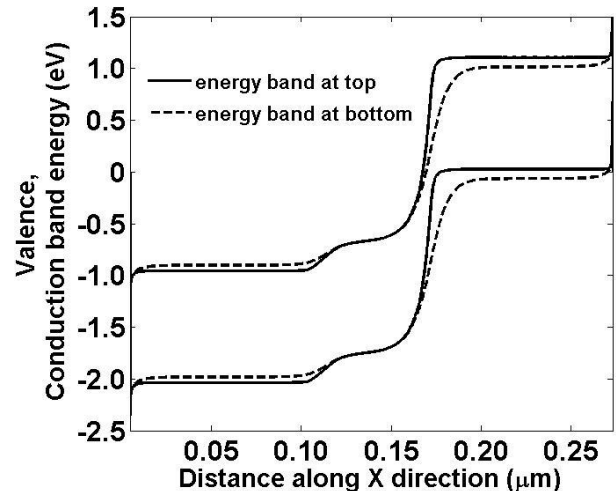

Fig. 5. Energy band diagrams in ON-state ($V_{GS}$ = $V_{DS}$ = 1.0 V) for the doping-less TFET along two horizontal cut-lines: one at the top and the other at the bottom of the silicon film (1 nm away from the oxide-silicon interface).

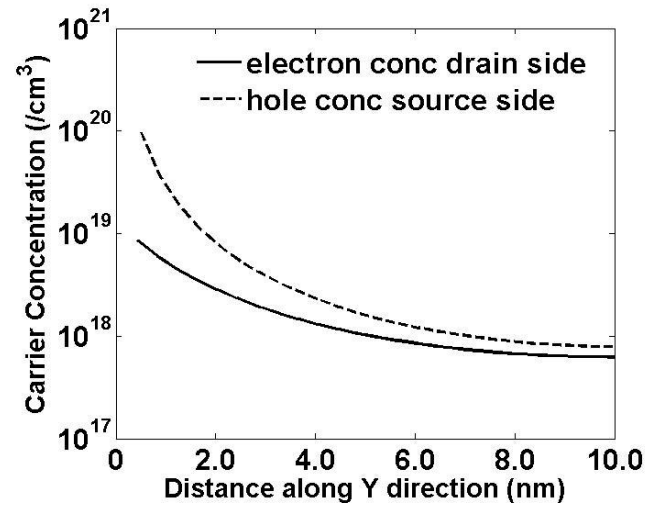

Fig. 6. Hole and electron concentrations in the source and drain regions, respectively, along the thickness of the silicon film for the doping-less TFET.

## III.  RESULTS AND DISCUSSION

The electron and hole concentrations for the doping-less TFET under thermal equilibrium ($V_{GS}$ = $V_{DS}$ = 0 V) and ON-state ($V_{GS}$ = $V_{DS}$  = 1.0 V) conditions are shown in Fig. 2. The induced thermal equilibrium hole and electron concentrations are similar to the source and drain dopings of the conventional TFET. What is to be noted is that in the doping-less TFET, the induced hole and electron carrier concentrations do not change significantly when the gate and drain voltages are applied. In Fig. 3, we have also shown the contour plots of the electron and hole carrier concentrations for $V_{DS}$ = 100 mV and 1.0 V with $V_{GS}$ fixed at 1.0 V. These plots help us in understanding how the electron and hole carrier concentrations vary across the device for different bias conditions.

The energy band diagrams of the doping-less TFET in the ON-state and OFF-state are shown in Fig. 4. In the ON-state, the valence band energy of the source side is aligned with the conduction band energy of the channel. This reduces the tunnel width in the ON-state increasing the tunneling probability. Looking at Fig. 4, it is clear that although the source and drain regions are created in the intrinsic silicon body without any doping, the carrier injection mechanism of this device is expected to be no different with respect to the conventional TFETs.

Fig. 5 shows the energy band diagrams for the doping-less TFET in ON-state ($V_{GS}$ = $V_{DS}$ = 1.0 V) along two horizontal cut-lines at the top and the bottom of the silicon film. We notice from Fig. 4 that the tunneling width on top of the silicon film is shorter compared to the tunneling width at the bottom of the silicon film. This difference in the tunneling width is due to the non-uniformity of the induced carrier concentrations in the source and drain regions of the doping-less TFET as shown in Fig. 6. In the conventional TFET, the source and drain regions are uniformly doped in the y-direction across the thickness of the silicon film. However, in the case of the doping-less TFET, the metal electrodes for inducing the free carriers are present only on the top side of the silicon film making the carrier concentrations asymmetric. On the other hand, if a FinFET like structure is used, the electrodes for inducing the charge carriers in the doping-less TFET can be present on both sides of the silicon film making the carrier concentrations identical on both sides of the silicon film.

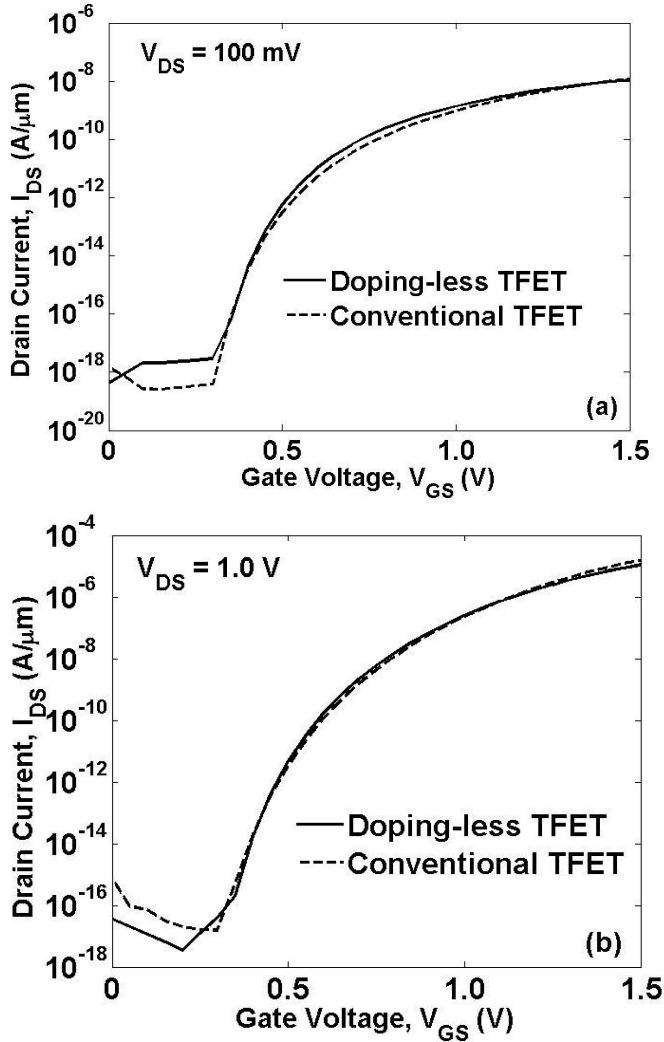

Fig. 7. Transfer characteristics of the doping-less TFET compared with that of the corresponding conventional TFET [6] for (a) $V_{DS}$ = 100 mV and (b) $V_{DS}$ = 1.0 V.

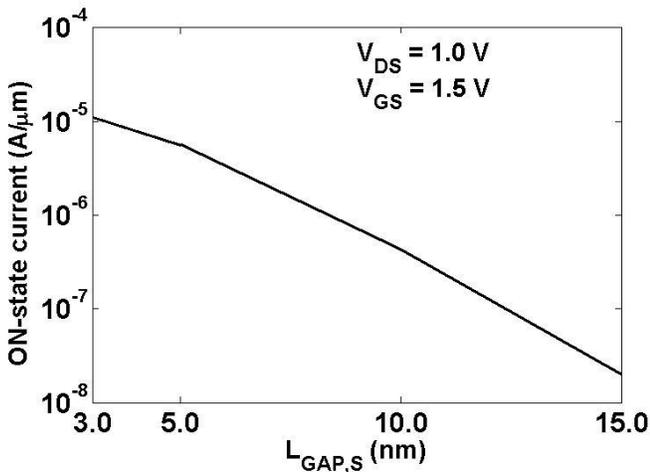

Fig. 8. ON-state current of the doping-less TFET versus the spacer thickness on the source side $L_{GAP,S}$.

The transfer characteristics of the doping-less TFET are shown in Fig. 7 and they follow a trend similar to that of a conventional TFET. To make sure that computational noise does not affect the accuracy of the calculated $I_{off}$, a very fine mesh is used in our simulations across the region where tunneling takes place. For $V_{DS}$ = 1.0 V and $V_{GS}$ = 0 V, the OFF-state current of the doping-less TFET is as low as $1 \times 10^{-17}$ A/μm. For $V_{DS}$ = 1.0 V and $V_{GS}$ = 1.5 V, the ON-state current of the doping-less TFET is ~$1.1 \times 10^{-5}$ A/μm which is similar to that of the conventional TFET [6].

The ON-state current can, however, be improved if the spacer thickness on the source side ($L_{GAP,S}$) is reduced. The spacer thickness between the source and gate electrodes determines the closeness of the gate field to the tunneling path on the source side. $L_{GAP,S}$ is an important parameter in determining the tunneling probability and hence the ON-state current and should be chosen carefully. Thin spacer oxides are commonly used in TFET fabrication and may potentially affect the device performance due to parasitic capacitance [20,38]. However, the advantages of a doping-less TFET such as immunity to random dopant fluctuations and lower thermal budgets make this device worth exploring.

Fig. 8 shows the ON-state current variation versus the spacer thickness on the source side ($L_{GAP,S}$) for the doping-less TFET. It is seen that the ON-state current reduces as $L_{GAP,S}$ is increased from 3 nm to 15 nm. An increase in $L_{GAP,S}$ reduces (i) the abruptness of the source-channel junction gradient leading to an increase in the tunneling width and (ii) moves the source-channel tunneling path away from the gate field reducing the tunneling efficiency. Therefore, $L_{GAP,S}$ should be as small as possible to get a good ON-state current. To reduce the dependence of ON-state current on process induced variations in $L_{GAP,S}$, the spacer oxide deposition needs to be done carefully using techniques such as atomic layer deposition (ALD) [39]. In conventional TFETs with doped junctions, it is important to have a sharp lateral source doping profile to enhance the transport of tunneling carriers [40]. However, to achieve a sharp doping profile, the process conditions need to be controlled very carefully using MBE, plasma implantation or laser annealing techniques [41]. On the other hand, in the case of doping-less TFET, it appears from Fig. 2 that steepness in the carrier profiles on the source side can be achieved by controlling the sidewall oxide spacer thickness. This property of the doping-less TFET could be exploited for improving the tunneling efficiency across the source-channel junction.

The output characteristics of the doping-less TFET are shown in Fig. 9. We observe that the doping-less TFET exhibits clear exponential and saturation regions of operation. The saturation region in the output characteristics is due to the tunneling width becoming progressively less dependent on $V_{DS}$ as the drain voltage is increased.

The average subthreshold swing for the doping-less TFET and the conventional TFET is calculated using [6]:

$$S = \frac{V_t - V_{off}}{log I_{vt} - log I_{off}}$$

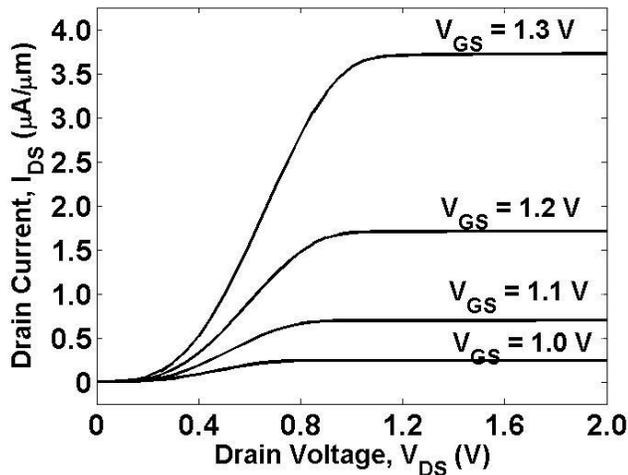

Fig. 9. Output characteristics of the doping-less TFET with saturation behavior similar to that of a MOSFET.

where the gate voltage at which the drain current becomes $1\times10^{-7}$ A/µm is taken as the threshold voltage $V_t$, $V_{off}$ is the gate voltage at which the device is in OFF-state, $I_{vt}$ is the drain current at threshold and $I_{off}$ is the OFF-state current when the gate voltage is zero. From Fig. 6 (b), the average subthreshold swing is found to be ~ 100 mV/decade for the doping-less TFET and is same as for the conventional TFET [6].

An increase in the ON-state current and a reduction in the average subthreshold swing of the doping-less TFET could be realized by using strain, high-k dielectric, narrow bandgap materials such as germanium or heterostructures.

## IV. CONCLUSION

A detailed study of the doping-less TFET using charge plasma concept is reported using two-dimensional simulations. Our results demonstrate that although the source and drain regions are induced on an intrinsic silicon body without the need for any doping, the source-channel tunneling process in the doping-less TFET can be controlled by a gate voltage similar to that of a conventional TFET. Due to the absence of dopant atoms in the doping-less TFET, it is expected to be immune to random dopant fluctuations. Moreover, the fabrication of the doping-less TFET does not demand high thermal budgets for creating the source and drain opening up the possibility of realizing TFETs on other substrates such as single-crystal silicon-on-glass. The results presented in this paper are preliminary in terms of ON-state current and subthreshold slope. However, many of the techniques that are currently being studied to improve these two parameters could also be employed to enhance the performance of the doping-less TFET. Our results may provide the incentive for further exploration of this device.

## REFERENCES

[1] A. C. Seabaugh and Q. Zhang, "Low-voltage tunnel transistors for beyond CMOS logic," *Proc. IEEE*, vol. 98, no. 12, pp. 2095–2110, Dec. 2010.

[2] A. M. Ionescu and H. Riel, "Tunnel field-effect transistors as energy efficient electronic switches," *Nature*, vol. 479, no. 7373, pp. 329–337, Nov. 2011.

[3] A. M. Ionescu, L. De Michielis, N. Dagtekin, G. Salvatore, J. Cao, A. Rusu, and S. Bartsch, "Ultra low power: Emerging devices and their benefits for integrated circuits," in *IEDM Tech. Dig.*, 2011, pp.16.1.1–16.1.4.

[4] K. K. Bhuwalka, S. Sedlmaier, A. K. Ludsteck, C. Tolksdorf, J. Schulze, and I. Eisele, "Vertical Tunnel Field-Effect Transistor," *IEEE Trans. Electron Devices*, vol.51. no.2, pp.279-281, February 2004.

[5] W. Y. Choi, B. G. Park, J. D. Lee and T. J. K. Liu, "Tunneling Field-effect Transistors (TFETs) with Subthreshold Swing (SS) Less Than 60 mV/dec," *IEEE Electron Device Letts.*,vol.28, no.8, pp.743-745, Aug. 2007.

[6] K.Boucart and A.M.Ionescu,"Double Gate Tunnel FET with High-k Gate Dielectric,"*IEEE Trans. Electron Devices*,vol.54, no.7,pp. 1725-1733, Jul. 2007.

[7] P. F. Wang, K. Hilsenbeck, T. Nirschl, M. Oswald, C. Stepper, M. Weis, D. Schmitt-Landsiedel and W. Hansch, "Complementary Tunneling Transistor for Low Power Application," *Sold-state Electronics*, vol.48 no.12, pp.2281-2286, Dec 2004.

[8] B. Ghosh and M. W. Akram, "Junctionless Tunnel Field Effect Transistor," *IEEE Electron Device Lett.*,vol.34, No.5, pp.584-586, May 2013.

[9] K. K. Bhuwalka, J. Schulze and I. Eisele, "Scaling the vertical tunnel FET with tunnel bandgap modulation and gate work function engineering,. *IEEE Trans. Electron Devices*, vol. 52, no. 5, pp. 909-917, 2005.

[10] Th. Nirschl, St. Henzler, J. Fischer, M. Fulde, A. Bargagli-Stoffi, M. Sterkel, J. Sedlmeir, C. Weber, R. Heinrich, U. Schaper, J. Einfeld, R. Neubert, U. Feldmann, K. Stahrenberg, E. Ruderer, G. Georgakos, A. Huber, R. Kakoschke, W. Hansch, D. Schmitt-Landsiedel, "Scaling properties of the tunneling field effect transistor (TFET): Device and circuit," *Solid-State Electronics*, Vol.50, pp.44-51, 2006.

[11] S.Saurabh and M.J. Kumar, "Impact of Strain on Drain Current and Threshold Voltage of Nanoscale Double Gate Tunnel Field Effect Transistor," *Japanese Journal of Appl. Phys.*, vol. 48, pt. 1, no. 6, article no.064503, June 2009.

[12] A. S. Verhulst, W. G. Vandenberghe, K. Maex, S. De Gendt, and M. Heyns and G. Groeseneken,"Complementary Silicon-Based Heterostructure Tunnel-FETs With High Tunnel Rates," *IEEE Electron Device Lett.*, vol.29, no.12, pp.1398-1401, Dec. 2008.

[13] T. Krishnamohan, D. Kim, C. D. Nguyen, C. Jungemann, Y. Nishi and K. C. Saraswat, "High-mobility Low Band-to-band Tunneling Strained-germanium Double-gate Heterostructure FETs: Simulations," *IEEE Trans. Electron Devices*, vol.53, no.5, pp.1000-1009, May 2006.

[14] Z. X. Chen, H. Y. Yu, N. Singh, N. S. Shen, R. D. Sayanthan, G. Q. Lo and D. L. Kwong, "Demonstration of Tunneling FETs Based on Highly Scalable Vertical Silicon Nanowires," *IEEE Electron Device Lett.*, vol.30, no.7, pp.754-756, July 2009.

[15] M.-H. Chiang, J.-N. Lin, K. Kim and C.-T. Chuang, "Random dopant fluctuation in limited-width FinFET technologies," *IEEE Tran. Electron Devices*, vol.54, no.8, pp. 2055-2060,Aug 2007.

[16] N. Damrongplasit, C. Shin, S. H. Kim, R. A. Vega and T. J. K. Liu, "Study of Random Dopant Fluctuation Effects in Germanium-Source Tunnel FETs," *IEEE Trans. Electron Devices*, vol.58, no.10, pp.3541-3548, October 2011.

[17] N. Damrongplasit, S. H. Kim and T. J. K. Liu, "Study of Random Dopant Fluctuation Induced Variability in the Raised-Ge-Source TFET," *IEEE Electron Device Letts.*, Vol.34, no.2, pp.184-186, February 2013.

[18] G. Leung, C. O. Chui, "Stochastic Variability in Silicon Double-Gate Lateral Tunnel Field-Effect Transistors," *IEEE Trans. Electron Devices*, vol.60, no.1, pp.84 - 91, January 2013.

[19] K. Boucart, A. M. Ionescu, and W. Riess, "A simulation-based study of sensitivity to parameter fluctuations of silicon tunnel FETs," in *Proc. ESSDERC*, 2010, pp. 345–348.

[20] C. L. Royer and F. Mayer, "Exhaustive Experimental Study of Tunnel Field Effect Transistors (TFETs): From Materials to Architecture," *10th Inter. Conf. on Ultimate Integration of Silicon*, 2009, pp.53-56.

[21] R.Jhaveri, V. Nagavarapu and J. C. S. Woo, "Effect of Pocket Doping and Annealing Schemes on the Source-Pocket Tunnel Field-Effect Transistor," *IEEE Trans. on Electron Devices*, Vol. 58, no.1, pp.80-86, January 2011.

[22] D. Leonelli, A. Vandooren, R. Rooyackers, S. D. Gendt, M. M. Heyns, G. Groeseneken, "Optimization of tunnel FETs: Impact of gate oxide

thickness, implantation and annealing conditions," in Proc. Solid-State Device Research Conference (ESSDERC), 2010, pp.170-173.

[23] B. Rajasekharan, R. J. E. Hueting, C. Salm, T. van Hemert, R. A. M. Wolters and J. Schmitz, "Fabrication and characterization of the charge-plasma diode," *IEEE Electron Device Lett.,* vol.31, no.6, pp.528-530, June 2010.

[24] R.J.E. Hueting, B. Rajasekharan, C. Salm and J. Schmitz, "Charge Plasma P-N Diode," *IEEE Electron Device Lett.,* vol.29, no.12, pp.1367-1368, Dec. 2008.

[25] M. J. Kumar and K. Nadda, "Bipolar Charge Plasma Transistor: A Novel Three Terminal Device," *IEEE Trans.. Electron Devices,* vol.59, no.4, pp.962-967, April 2012.

[26] K. Nadda and M. J. Kumar, "Schottky Collector Bipolar Transistor without Impurity Doped Emitter and Base: Design and Performance", Accepted in *IEEE Trans. Electron Devices,* Vol.60, 2013.

[27] L. Lattanzio, L. De Michielis and A. M. Ionescu, "Complementary Germanium Electron–Hole Bilayer Tunnel FET for Sub-0.5-V Operation," *IEEE Electron Device Lett.,*Vol.33, no.2, pp.167-169, February 2012.

[28] L. Lattanzio, A. Biswas, L. D. Michielis, and A. M. Ionescu, "A tunneling field-effect transistor exploiting internally combined band-toband and barrier tunneling mechanisms," *Appl. Phys. Lett.,* Vol.98, Article No.123504, 2011.

[29] W. H. Teh, A. Trigg, C. H. Tung, R. Kumar, N. Balasubramanian, and D. L. Kwong, "200 mm wafer-scale epitaxial transfer of single crystal Si on glass by anodic bonding of silicon-on-insulator wafers," *Appl. Phys. Letts.,* vol.87, article no.073107, 2005.

[30] H. S. Kima, R. H. Blick, D. M. Kim and C. B. Eom, "Bonding silicon-on-insulator to glass wafers for integrated bio-electronic circuits," *Appl. Phys. Letts.,* vol.85, no.12, pp.2370-2372, September 2004.

[31] S. Saurabh and M. J. Kumar, "Estimation and Compensation of Process Induced Variations in Nanoscale Tunnel Field Effect Transistors (TFETs) for Improved Reliability," *IEEE Trans. on Device and Materials Reliability*, vol.10, no.9, pp.390-395, Sept. 2010.

[32] S. Saurabh and M. J. Kumar, "Investigation of the Novel Attributes of a Dual Material Gate Nanoscale Tunnel Field Effect Transistor," *IEEE Trans. on Electron Devices*, vol.58, no.2, pp.404-410, Feb. 2011.

[33] Y. Omura, S. Horiguchi, M. Tabe and K. Kishi, "Quantum-Mechanical Effects on the Threshold Voltage of Ultrathin-SOI nMOSFETs," *IEEE Trans. on Electron Devices*, vol.14, no.12, pp.569-571, December 1993.

[34] D. R. Lide, *CRC Handbook on Chemistry and Physics*, 89th ed. New York, NY, USA: Taylor & Francis, 2008, pp.12-114.

[35] J. Shi, D. Kojima and M. Hashimoto, "The interaction between platinum films and silicon substrates: Effects of substrate bias during sputtering deposition", *J. Appl. Phys.,* Vol.88, pp.1679-1683, 2000.

[36] *ATLAS Device Simulation Software*, Silvaco Int., Santa Clara, CA, 2012.

[37] A. Schenk, "Finite-temperature full random-phase approximation model of band gap narrowing for silicon device simulation", *J. Appl. Phys.,* Vol.84, pp.3684-3695, 1998.

[38] H. G. Virani, R. B. R. Adari and A. Kottantharayil, "Dual-k Spacer Device Architecture for the Improvement of Performance of Silicon n-Channel Tunnel FETs," *IEEE Trans. Electron Devices,* vol.57, no.10, pp.2410-2417, October 2010.

[39] G.-W. Lee, H.-D. Lee, K.-Y. Lim, Y. S. Kim, H.-S. Yang, G.-S. Cho, S.-K. Park and S.-J. Hong, "Characterization of Polymetal Gate Transistors With Low-Temperature Atomic-Layer-Deposition-Grown Oxide Spacer," *IEEE Electron Device Lett.,* Vol. 30, N0. 2, pp.181-183, February 2009.

[40] H.-Y. Chang, S. Chopra, B. Adams, J. Li, S. Sharma, Y. Kim, S. Moffatt, J. C. S. Woo, "Improved subthreshold characteristics in tunnel field-effect transistors using shallow junction technologies", *Solid-State Electron.,* Vol.80, pp.59-62, 2013.

[41] D. Leonelli, A. Vandooren, R. Rooyackers, S. De Gendt, M.M. Heyns, G. Groeseneken, "Drive current enhancement in p-tunnel FETs by optimization of the process conditions", *Solid-State Electron.,* Vol.65–66, pp.28-32, 2011.

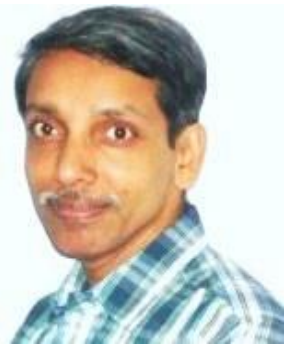

**M. Jagadesh Kumar** is the Chair Professor of the NXP (Philips) (currently, NXP Semiconductors India Pvt. Ltd.) established at IIT Delhi by Philips Semiconductors, The Netherlands. He is also a Principal Investigator of the Nano-scale Research Facility at IIT Delhi. He is an Editor of the IEEE TRANSACTIONS ON ELECTRON DEVICES. For more details on Dr. Jagadesh Kumar, please visit http://web.iitd.ac.in/~mamidala

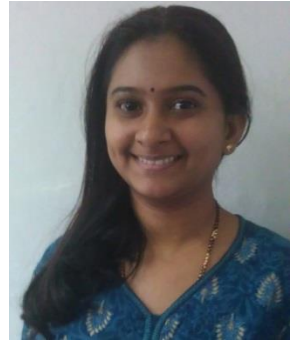

**Sindhu Janardhanan** received her B. Tech degree in electronics and communication engineering from Cochin University of Science and Technology, Cochin, India and M.Tech degree in very large scale integration design from Amrita Vishwa Vidyapeetham, Coimbatore, India. She is working towards a Ph.D degree in Department of Electrical Engineering, Indian Institute of Technology, Delhi, India. Her research interests include nanoscale devices and modeling.